\documentstyle[pre,aps,eqsecnum]{revtex}
\begin{document}

\large
\draft
\title{Notes about Passive Scalar in \\
Large-Scale Velocity Field}
\author{I. Kolokolov$^{a,e}$, V. Lebedev$^{b,c}$, and M. Stepanov$^{d,e}$.}
\address{$^a$ Budker Institute of Nuclear Physics,
Novosibirsk 630090, Russia; \\
$^b$ Landau Institute for Theoretical Physics, RAS, \\
Kosygina 2, Moscow 117940, Russia; \\
$^c$ Department of Physics of Complex Systems,
Weizmann Institute of Science, \\ Rehovot 76100, Israel; \\
$^d$ Institute of Automation and Electrometry, RAS, \\
Novosibirsk 630090, Russia; \\
$^e$ Novosibirsk State University, Novosibirsk, 630090, Russia.}
\date{\today}
\maketitle

\begin{abstract} \large
  We consider advection of a passive scalar $\theta(t,{\bbox r})$ by
  an incompressible large-scale turbulent flow. In the framework of
  the Kraichnan model the whole PDF's (probability distribution
  functions) for the single-point statistics of $\theta$ and for the
  passive scalar difference $\theta({\bbox r}_1)-\theta({\bbox r}_2)$
  (for separations ${\bbox r}_1-{\bbox r}_2$ lying in the convective
  interval) are found.
\end{abstract}

\section*{Introduction}

We treat advection of a passive scalar field $\theta(t,{\bbox r})$ by an
incompressible turbulent flow, the role of the scalar can be played by
temperature or by pollutants density. The velocity field is assumed to
contain motions from some interval of scales restricted from below by $L_v$.
A steady situation with a permanent random supply of the passive scalar is
considered. We wish to establish statistics of the passive scalar $\theta$
for scales that are less than both the scale $L_v$ and the pumping scale $L$,
and larger than the diffusion scale $r_{\rm dif}$  (for definiteness we
assume that $L<L_v$). Such convective interval of scales exists if the Peclet
number ${\rm Pe}=L/r_{\rm dif}$ is large enough, we will assume the
condition. Since all scales from the convective interval are assumed to be
smaller than $L_v$ we will say about advection by a large-scale turbulent
flow. The problem is of physical interest for the dimensionalities
$d=2,3$, but formally it can be treated for an arbitrary dimensionality $d$
of space. Below we will treat $d$ as a parameter. Particularly, all the
expressions will be true for a space of large dimensionality $d$.

Description of a small-scale statistics of a passive scalar advected by a
large-scale solenoidal velocity field is a special problem in turbulence
theory. This problem was treated consistently from the very beginning and
some rigorous results have been obtained which is quite unusual for a
turbulence problem. Batchelor \cite{Bat} examined the case of external
velocity field being so slow that it does not change during the time of the
spectral transfer of the scalar from the external scale to the diffusion
scale. Then Kraichnan \cite{Kra} obtained plenty of results in the opposite
limit of a velocity field delta-correlated in time. The pair correlation
function of the passive scalar
$\langle\theta({\bbox r})\theta({\bbox 0})\rangle$ was found to be
proportional to the logarithm $\ln(L/r)$ and the pair
correlation function of the passive scalar difference
$\langle[\theta({\bbox r})-\theta({\bbox 0})]^2\rangle$ was found to be
proportional to $\ln(r/r_{\rm dif})$ in both cases. The assertions are really
correct for any temporal statistics of the velocity field
\cite{94SS,95CFKLa}. Thus we are dealing with the logarithmic case which is
substantially simpler than cases with power-like correlation functions
usually encountered in turbulence problems \cite{53Bat,MY,Frish}.

Now about high-order correlation functions of the passive scalar. As long as
all distances between the points are much less than $L$, the $2n$-point
correlation functions of $\theta$ are given by their reducible parts (that is
are expressed via products of the pair correlation function) until
$n\sim\ln(L/r)$ where $r$ is either the smallest distance between the
points or $r_{\rm dif}$ depending on what is larger \cite{95CFKLa}. The
reason for such Wick decoupling is simply the fact that reducible parts
contain more logarithmic factors (which are considered as large ones) than
non-reducible parts do. Consistent calculations of the fourth-order
correlation function of the passive scalar at $d=2$ \cite{95BCKL} confirm the
assertion. Therefore e.g. the single-point PDF of $\theta$ has a Gaussian
core (that describes the first moments with $n<\ln{\rm Pe}$) and a
non-Gaussian tail (that describes moments with $n>\ln{\rm Pe}$). The tail
appears to be exponential (see \cite{94SS,95CFKLa}). The same is true for the
passive scalar difference $\Delta\theta=\theta({\bbox r})-\theta({\bbox 0})$
where instead of $\ln{\rm Pe}$ we should take $\ln(r/r_{\rm dif})$. The tails
do not depend on $\ln{\rm Pe}$ or on $\ln(r/r_{\rm dif})$ and contain only
coefficients depending on the statistics of the advecting velocity.

Correlation functions of the passive scalar can be written as averages of
integrals of the pumping along Lagrangian trajectories (see e.g.
\cite{94FL}). For example, the pair correlation function
$\langle\theta({\bbox r})\theta({\bbox 0})\rangle$ is proportional to
an average time needed for two points moving along Lagrangian trajectories to
run from the distance $r$ to the distance $L$. Generally, correlation
functions of the passive scalar are determined by the spectral transfer that
is by an evolution of Lagrangian separations up to the scale $L$. For the
large-scale velocity field Lagrangian dynamics is determined by the
stretching matrix $\sigma_{\alpha\beta} =\nabla_\beta v_\alpha$ and,
consequently, the statistics of the matrix determines correlation functions
of the passive scalar. For example, the coefficient at the logarithm in the
pair correlation function of the passive scalar is $P_2/\bar\lambda$
\cite{Bat,Kra,94SS,95CFKLa} where $P_2$ is the pumping rate of
$\theta^2$ and $\bar\lambda$ is Lyapunov exponent that is the average
of the largest eigen value of the matrix $\hat\sigma$. The
coefficients in the exponential tails are more sensitive to the
statistics of $\hat\sigma$, particurlaly they depend on the
dimensionless parameter $\bar\lambda\tau$ \cite{95CFKLa} where $\tau$
is the correlation time of $\hat\sigma$. The motion of the fluid particles
in the random velocity field resembles in some respects random walks, but one
should remember that correlation lengths of both the advecting
velocity and of the pumping are much larger than scales from the
convective interval we are interested in. Thus the situation is
opposite to one usually occurring in solid state physics, where e.g.
random potential is shortly correlated in space.

Since really $\ln(L/r)$ is not very large it is of interest to find
the whole PDF's for the single-point statistics of $\theta$ and for
the passive scalar difference $\Delta\theta$. It is possible to do for the
Kraichnan short-correlated case $\bar\lambda\tau\ll1$ when the statistics of
$\hat\sigma$ can be regarded to be Gaussian. An attempt to do this was made
in \cite{94CGK,96FKLM} in terms of the statistics of the main eigen value of
the matrix $\hat\sigma$.  Unfortunately the scheme works only for the
dimensionality $d=2$ where the matrix $\hat\sigma$ has a single eigen value.
This fact was noted in the work \cite{98BGK} where also the correct
coefficient in the exponential tails for an arbitrary dimensionality of space
$d$ was found. Here, we develop a scheme enabling to obtain the whole PDF's
for arbitrary $d$. The scheme is also interesting from the methodical point
of view. For example, its modification enables one to calculate the
statistics of local dissipation \cite{98GK}.

The paper is organized as follows. In Section \ref{sec:gen} we
find a path integral representation for the simultaneous statistics of
the passive scalar. In Section \ref{sec:fun} we analyze the generating
functional for correlation functions of the passive scalar in the convective
interval of scales. Using different approaches we obtain the functional and
establish the applicability conditions of our consideration. In Section
\ref{sec:pdf} we find explicit expressions for the single-point PDF and for
the PDF of the passive scalar difference. In Conclusion we shortly discuss the
obtained results.

\section{General Relations}
\label{sec:gen}

The dynamics of the passive scalar $\theta$ advected by the velocity
field ${\bbox v}$ is described by the equation
\begin{equation}
  \partial_t\theta+{\bbox v}\nabla\theta
    -\kappa\nabla^2\theta=\phi \,.
  \label{ga1}
\end{equation}
Here, the term with the velocity ${\bbox v}$ describes the advection
of the passive scalar, the next term is diffusive ($\kappa$ is the
diffusion coefficient) and $\phi$ describes a pumping source of the
passive scalar. Both ${\bbox v}(t,{\bbox r})$ and $\phi(t,{\bbox r})$
are supposed to be random functions of $t$ and ${\bbox r}$. We regard
the statistics of the velocity and of the source to be independent.
Therefore, all correlation functions of $\theta$ are to be treated as
averages over both statistics.

\subsection{Simultaneous Statistics}
\label{simsta}

The source $\phi$ is believed to possess a Gaussian statistics and to
be $\delta$-correlated in time. The statistics is entirely
characterized by the pair correlation function
\begin{equation}
  \langle\phi(t_1,{\bbox r}_1)\phi(t_2,{\bbox r}_2)\rangle
    =\delta(t_1-t_2)\chi\left(|{\bbox r}_1-{\bbox r}_1|\right) \,,
  \label{gaa1}
\end{equation}
where we assumed that the pumping is isotropic. The function $\chi(r)$ is
accepted to have a characteristic scale $L$ which is the pumping length. We
will be interested in the statistics of the passive scalar on scales much
smaller than $L$.

Simultaneous correlation functions of the passive scalar $\theta$ can be
represented as coefficients of the expansion over $y$ of the generating
functional
\begin{equation}
  {\cal Z}(y)=\left\langle\exp\left\{{\rm i}y
   \int {\rm d}{\bbox r}\, \beta({\bbox r})
    \theta(0,{\bbox r})\right\}\right\rangle \,,
  \label{gaa3}
\end{equation}
where $\beta$ is a function of coordinates and angular brackets
designate averaging over both the statistics of the pumping $\phi$
and over the statistics of the velocity ${\bbox v}$. The generating
functional ${\cal Z}(y)$ contains the entire information about the
simultaneous statistics of the passive scalar $\vartheta$. Particularly,
knowing ${\cal Z}(y)$ one can reconstruct the simultaneous PDF of the passive
scalar, the problem is discussed in Section \ref{sec:pdf}.

If characteristic scales of $\beta$ in (\ref{ga5}) are much larger than the
diffusion scale $r_{\rm dif}$ then it is possible to neglect diffusivity at
treating the generating functional (\ref{gaa3}). Then the left-hand side of
the equation (\ref{ga1}) describes the simple advection and it is reasonable
to consider a solution of the equation in terms of Lagrangian trajectories
${\bbox\varrho}(t)$ introduced by the equation
\begin{equation}
  \partial_t{\bbox\varrho}=
    {\bbox v}(t,{\bbox\varrho}) \,.
  \label{ga2}
\end{equation}
We will label the trajectories by ${\bbox r}$, those are positions of
Lagrange particles at $t=0$: ${\bbox\varrho}(0,{\bbox r})={\bbox r}$.
Next, introducing $\tilde\theta(t,{\bbox r})=\theta(t,{\bbox\varrho})$
we rewrite the equation (\ref{ga1}) as $\partial_t\tilde\theta=\phi$, that
leads us to
\begin{equation}
  \theta(0,{\bbox r})=\int_{-\infty}^0\!\! {\rm d}t\,
    \phi(t,{\bbox\varrho}) \,.
  \label{ga3}
\end{equation}
Here we have taken into account that at $t=0$ the functions $\theta$
and $\tilde\theta$ coincide. Starting from (\ref{ga3}) and using Gaussianity
of the pumping statistics we can average explicitly the generating functional
(\ref{gaa3}) over the statistics. The result is
\begin{eqnarray}
  && {\cal Z}(y)=\left\langle\exp\left[
    -\frac{y^2}{2}\int_{-\infty}^0\!\! {\rm d}t\,U\right]
    \right\rangle  \,, \label{gaa5} \\
  && U=\int {\rm d}{\bbox r}_1 {\rm d}{\bbox r}_2\,
    \beta({\bbox r}_1)\beta({\bbox r}_2)
    \chi(|{\bbox\varrho}_1-{\bbox\varrho}_2|) \,, \label{ga5}
\end{eqnarray}
where angular brackets mean averaging over the statistics of the
velocity field only.

Being interested in the single-point statistics of $\theta$ we should
take $\beta({\bbox r})=\delta({\bbox r})$. But it is impossible since
we have neglected diffusivity. We will take
$\beta({\bbox r})=\delta_\Lambda({\bbox r})$ instead where the function
$\delta_\Lambda({\bbox r})$ tends to zero at $\Lambda r>1$ fast
enough and is normalized by the condition
\begin{eqnarray} &&
  \int d{\bbox r}\, \delta_\Lambda({\bbox r})=1 \,.
\nonumber \end{eqnarray}
Then the generating functional (\ref{gaa5}) will describe the
statistics of an object
\begin{eqnarray} &&
  \theta_{\Lambda}=\int{\rm d}{\bbox r}\,
    \delta_\Lambda({\bbox r})\theta({\bbox r}) \,,
\label{theta} \end{eqnarray}
smeared over a spot of the size $\Lambda^{-1}$. If $r_{\rm dif}\Lambda\ll1$
then the statistics of the object is not sensitive to diffusivity. From the
other hand if $\Lambda L\gg1$, then knowing correlation functions of
$\theta_{\Lambda}$, we can reconstruct single-point statistics due to the
logarithmic character of the correlation functions. To obtain single-point
correlation functions one should substitute simply
$\Lambda\to r_{\rm dif}^{-1}$ into the correlation functions of
$\theta_{\Lambda}$. The above inequalities $\Lambda r_{\rm dif}\ll1$ and
$\Lambda L\gg1$ are compatible because of ${\rm Pe}\gg1$. If we are
interested in the statistics of the passive scalar differences in points with
a separation ${\bbox r}_0$ (where $r_0\gg r_{\rm dif}$) then instead of
$\delta_\Lambda({\bbox r})$ we should take
\begin{eqnarray} &&
  \beta({\bbox r})= \delta_\Lambda({\bbox r}-{\bbox r}_0/2)
    -\delta_\Lambda({\bbox r}+{\bbox r}_0/2) \,.
\label{difb} \end{eqnarray}
Then the  generating functional (\ref{gaa5}) will describe the
statistics of an object
\begin{eqnarray} &&
  \Delta\theta_{\Lambda}=\theta_{\Lambda}({\bbox r}_0/2)
     -\theta_{\Lambda}(-{\bbox r}_0/2) \,.
\label{diff} \end{eqnarray}
Again, correlation functions of the passive scalar differences can be
found from correlation functions of $\Delta\theta_{\Lambda}$ after the
substitution $\Lambda\to r_{\rm dif}^{-1}$.

\subsection{Path Integral}

Below, we treat advection of the passive scalar by a large-scale velocity
field, that is we assume that the velocity correlation length $L_v$ is larger
than the scales from the convective interval. Then for the scales one can
expand the difference
\begin{eqnarray} &&
  v_\alpha({\bbox r}_1)- v_\alpha({\bbox r}_2)
    =\sigma_{\alpha\beta}(t)(r_{1\beta}-r_{2\beta}) \,, \quad
      \sigma_{\alpha\beta}=\nabla_\beta v_\alpha \,.
\label{ga8} \end{eqnarray}
Here $\sigma_{\alpha\beta}(t)$ can be treated as an ${\bbox
r}$-independent matrix field. Then the equation (\ref{ga2}) leads to
\begin{equation}
  \partial_t({\varrho}_{1,\alpha}-{\varrho}_{2,\alpha})=
    \sigma_{\alpha\beta}(t)({\varrho}_{1,\beta}-{\varrho}_{2,\beta})\,.
  \label{gaa4}
\end{equation}
A formal solution of the equation (\ref{gaa4}) is
\begin{eqnarray} &&
  {\varrho}_{1,\alpha}-{\varrho}_{2,\alpha}
    =W_{\alpha\beta}(r_{1,\beta}-r_{2,\beta}) \,,
\nonumber \\ &&
  \partial_t \hat W =\hat\sigma\hat W \,, \quad \hat W={\cal
    T}\exp\left(-\int_t^0 \!\! {\rm d}t\,\hat\sigma\right) \,,
\label{ga7} \end{eqnarray}
where ${\cal T}$ means an antichronological ordering.  Note that
${\rm Det\,}{\hat W}=1$, those property is a consequence of ${\rm
Tr}\,\hat\sigma=0$ and the initial condition $\hat W=1$ at $t=0$. The
Lagrangian difference in (\ref{ga5}) is now rewritten as
\begin{eqnarray}
  && |{\bbox\varrho}_1-{\bbox\varrho}_2|
    =\sqrt{(r_{1\alpha}-r_{2\alpha})
    B_{\alpha\beta}(r_{1\beta}-r_{2\beta})} \,, \quad
    \hat B=\hat W^T \hat W \,,
  \label{gaa7}
\end{eqnarray}
where the subscript $T$ designates a transposed matrix. Note that
${\rm Det\,}{\hat B}=1$ since ${\rm Det\,}{\hat W}=1$.

The generating functional ${\cal Z}(y)$ (\ref{gaa5}) can be
explicitly calculated in the Kraichnan case \cite{Kra} when the
statistics of the velocity is $\delta$-correlated in time. Then the
velocity statistics is Gaussian and is entirely determined by the
pair correlation function which in the convective interval is written as
\begin{eqnarray} &&
 \left\langle v_\alpha(t_1,{\bbox r}_1)
  v_\beta(t_2,{\bbox r}_2)\right\rangle=\delta(t_1-t_2)
   \left[{\cal V}_0\delta_{\alpha\beta}
    -{\cal K}_{\alpha\beta}({\bbox r}_1-{\bbox r}_2)\right] \,,
     \label{vel} \\ &&
    {\cal K}_{\alpha\beta}({\bbox r})=
   D(r^2\delta_{\alpha\beta}-r_\alpha r_\beta)
 +\frac{(d-1)D}{2}\delta_{\alpha\beta}\, r^2 \,.
\label{vel1} \end{eqnarray}
Here ${\cal V}_0$ is a huge ${\bbox r}$-independent constant and $D$
is a parameter characterizing the amplitude of the strain fluctuations. The
structure of (\ref{vel1}) is determined by the isotropy and space
homogeneouty accepted, and by the incompressibility condition
${\rm div}\,{\bbox v}=0$. Then the statistics of $\hat\sigma$ is Gaussian and
is determined by the pair correlation function which can be found from
(\ref{vel},\ref{vel1}):
\begin{equation}
 \left\langle\sigma_{\alpha\beta}(t_1)\sigma_{\mu\nu}(t_2)\right\rangle
  =D\left[(d+1)\delta_{\alpha\mu}\delta_{\beta\nu}
  -\delta_{\alpha\nu}\delta_{\beta\mu}
 -\delta_{\alpha\beta}\delta_{\mu\nu}\right]\delta(t_1-t_2) \,.
\label{ga9} \end{equation}
Note that the correlation function (\ref{ga9}) is
${\bbox r}$-independent, as it should be. We see from (\ref{ga9}) that the
parameter $D$ characterizes amplitudes of $\hat\sigma$ fluctuations.

Averaging over the statistics of $\hat\sigma$ can be substituted by
the path integral over unimodular matrices $\hat W(t)$ with a weight
$\exp\left(i{\cal I}\right)$. The effective action ${\cal I}=\int{\rm
d}t\,{\cal L}_0$ is determined by (\ref{ga9}):
\begin{equation}
  {\rm i}{\cal L}_0=-\frac{1}{2d(d+2)D}
    \left[(d+1){\rm Tr}(\hat\sigma^T\hat\sigma)
      +{\rm Tr}\hat\sigma^2\right] \,.
\label{ga17} \end{equation}
Then the generating functional (\ref{ga5}) can be rewritten as the
following functional integral over unimodular matrices
\begin{eqnarray}
  && {\cal Z}(y)=\int {\cal D}\hat W \,
    \exp\left[\int_{-\infty}^0 \!\! {\rm d}t\, \
     \left( {\rm i}{\cal L}_0-\frac{y^2}{2}U\right)\right] \,,
    \label{ga16} \\
  && U = \int {\rm d}{\bbox r}_1 {\rm d}{\bbox r}_2 \,
    \beta({\bbox r}_1)\beta({\bbox r}_2)
     \chi\left[\sqrt{(r_{1\alpha}-r_{2\alpha})
    B_{\alpha\beta}(r_{1\beta}-r_{2\beta})}\right] \,.
    \label{ga18}
\end{eqnarray}
Here, we should substitute $\hat\sigma=\partial_t \hat W(\hat W)^{-1}$
and remember about the boundary condition $\hat W=1$ at $t=0$.

Some words about the `potential' $U$ (\ref{ga5}) figuring in (\ref{ga18}).
The characteristic value of ${\bbox r}_1-{\bbox r}_2$ in the integral
(\ref{ga5}) is of order $\Lambda^{-1}$ for
$\beta({\bbox r})=\delta_\Lambda({\bbox r})$. Since we assume $\Lambda L\gg1$
then for a single-point statistics $U\approx P_2$, where $P_2=\chi(0)$, if
$B$ is not very large.  Particularly it is correct at moderate times $|t|$
since $\hat B=\hat 1$ at $t=0$. With increasing $|t|$ the argument of $\chi$
in (\ref{ga18}) grows and $U$ tends to zero when the argument of $\chi$
becomes greater than $L$.  For the passive scalar difference when $\beta$ is
determined by (\ref{difb}) the situation is a bit more complicated. Then $U$
is a difference of two contributions. The first contribution behaves like for
a single-point statistics. The second contribution contains $\chi$ with the
argument determined by ${\bbox r}_1-{\bbox r}_2\approx\pm{\bbox r}_0$. Then
at $t=0$ the meaning of the second contribution is determined again by $P_2$
but it turns to zero at increasing $|t|$ earlier than the first contribution.

Thus we reduced our problem to the path integral (\ref{ga16}) that is
to the quantum mechanics with $d^2-1$ degrees of freedom. Nevertheless to
solve the problem we should perform an additional reduction of the degrees of
freedom. The conventional way to do this is passing to eigen values, say, of
the matrix $\hat B$ figuring in (\ref{ga18}) (see e.g. \cite{mehta}) and
excluding angular degrees of freedom. Just this way was used by Bernard,
Gawedzki and Kupiainen in \cite{98BGK}. Then the authors using known facts
about the quantum mechanics associated with the eigenvalues (see e.g.
\cite{OlPe}) have found the coefficient in the exponential tail in the
single-point PDF of $\theta$. Unfortunately this way is not very convenient
to find the whole PDF. To do this we will use a special representation of the
matrix $\hat W$ in the spirit of the nonlinear substitution introduced by
Kolokolov \cite{Kol}. That is a subject of the next subsection.

\subsection{Choice of Parametrization}

To examine the generating functional ${\cal Z}(y)$ we will use
a mixed rotational-triangle parametrization
\begin{equation}
  \hat W=\hat R \hat T \,, \quad \hat B=\hat T^T\hat T \,,
   \label{gy1}
\end{equation}
where $\hat R$ is an orthogonal matrix and $\hat T$ is a triangular
matrix: $T_{ij}=0$ at $i>j$. The parametrization (\ref{gy1}) is the
direct generalization of the $2d$ substitution suggested in
\cite{98CFK}. Note that ${\rm Det}\,\hat T=1$ since ${\rm Det}\,\hat
W=1$. Note also that the matrix $\hat B$ introduced by (\ref{gaa7})
does not depend on $\hat R$ as is seen from (\ref{gy1}). That is a
motivation to exclude the matrix $\hat R$ from the consideration
performing the integration over the corresponding degrees of freedom
in the path integral (\ref{ga16}). A jacobian appears at the
integration. To avoid an explicit calculation of the jacobian (which
needs a discretization over time and then an analysis of an infinite
matrix \cite{94CGK}) we will use an alternative procedure described below.

Let us examine the dynamics of the matrix $\hat T$. It is determined
by the equation
\begin{equation}
  \partial_t T_{ij}=\Sigma_{ii}T_{ij}
      +\sum\limits_{i<k\le j}(\Sigma_{ik}+\Sigma_{ki})T_{kj} \,,
\label{gy3} \end{equation}
following from (\ref{ga7},\ref{gy1}). Here we used the designation
\begin{equation}
  {\hat\Sigma}=\hat R^T\hat\sigma\hat R \,.
  \label{ga12}
\end{equation}
Next introducing the quantities
\begin{equation}
  T_{ii} = \exp(\rho_i)\,, \quad
    T_{ij} = \exp(\rho_i)\eta_{ij} \,, \quad {\rm if}\ i < j \,,
  \label{gy5}
\end{equation}
we rewrite the equation (\ref{gy3}) as
\begin{eqnarray} &&
   \partial_t \rho_i=\Sigma_{ii} \,,
\label{gy6} \\ &&
  \partial_t \eta_{ij}=(\Sigma_{ij}+\Sigma_{ji})
    \exp(\rho_j-\rho_i)
     +\sum\limits_{i<k<j}(\Sigma_{ik}+\Sigma_{ki})
      \exp(\rho_k-\rho_i)\eta_{kj} \,.
\label{gy7} \end{eqnarray}
Comparing (\ref{ga7}) with (\ref{gy1}) one can find the following
expression for $\hat A=\hat R^T\partial_t\hat R$
\begin{equation}
  A_{ij}=\Sigma_{ij} \quad {\rm if}\ i > j \qquad
    A_{ij}=-\Sigma_{ji} \quad {\rm if}\ i < j \,.
  \label{gy2}
\end{equation}

One can easily check that the irreducible pair correlation function of
$\Sigma_{ij}$ has the same form as for $\sigma_{ij}$ that is
determined by (\ref{ga9}):
\begin{eqnarray} &&
  \langle\Sigma_{ij}(t_1)\Sigma_{mn}(t_2)\rangle
    =D[(d+1)\delta_{im}\delta_{jn}
     -\delta_{in}\delta_{jm}
       -\delta_{ij}\delta_{mn}]\delta(t_1-t_2) \,.
\label{sig} \end{eqnarray}
Besides, the average value of $\Sigma_{ij}$ is nonzero \cite{94CGK}:
\begin{equation}
  \langle \Sigma_{ij}\rangle =
    - D\frac{d(d-2i+1)}2 \delta_{ij} \,.
  \label{gy4}
\end{equation}
Nonzero averages of $\Sigma_{ij}$ are related to Lyapunov exponents
(not only the first one) \cite{83Dor} (for our model see
also \cite{KG}). To obtain (\ref{gy4}) one should first
solve the equation $\hat A=\hat R^T\partial_t\hat R$ for
$\hat R$ on a small interval $\tau$:

\begin{eqnarray} &&
  \hat R(t+\tau)\approx\hat R(t)\left[1
    +\int_t^{t+\tau}\!\!\!{\rm d}t'\,\hat A(t')\right] \,.
\nonumber  \end{eqnarray}
Then with the same accuracy we get from the expression (\ref{ga12})
\begin{eqnarray} &&
  \hat\Sigma(t+\tau)\approx
    \hat R^T(t)\hat\sigma(t+\tau)\hat R(t)
      +\left[\hat\Sigma(t+\tau),
        \int_t^{t+\tau}\!\!\!{\rm d}t'\,\hat A(t')\right] \,.
\label{rox} \end{eqnarray}
The average value of $\hat\Sigma$ arises from the second term in the
right-hand side of (\ref{rox}). The explicit form of the average can
be found using
\begin{eqnarray} &&
 \left\langle\Sigma_{ij}(t+\tau)
  \int_t^{t+\tau}\!\!\!{\rm d}t'\,
   \Sigma_{mn}(t')\right\rangle=
    \frac{D}{2}[(d+1)\delta_{im}\delta_{jn}
     -\delta_{in}\delta_{jm}-\delta_{ij}\delta_{mn}] \,.
\label{rox1} \end{eqnarray}
Here we utilized Eq. (\ref{sig}) and substituted the integral
\begin{eqnarray}
\int_t^{t+\tau}\!\!\!{\rm d}t'\,\delta(t+\tau-t')
\nonumber \end{eqnarray}
by $1/2$. The reason is that really the correlation function of
$\hat\sigma$ has a finite correlation time and therefore $\delta(t)$
(representing this correlation function) should be substituted by a
narrow function symmetric under $t\to-t$. Then we will get $1/2$.
Expressing in (\ref{rox}) $\hat A$ via $\hat\Sigma$ from (\ref{gy2})
and calculating its average using (\ref{rox1}) we get the answer (\ref{gy4}).

The expressions (\ref{gy6},\ref{gy7},\ref{sig},\ref{gy4}) entirely
determine the stochastic dynamics of $\rho_i$ and $\eta_{ij}$. Using
the conventional approach \cite{61Wyl,73MSR,76Dom,76Jan,DP78} correlation
functions of these degrees of freedom can be described in terms of a path
integral over $\rho_i$, $\eta_{ij}$ and over auxiliary fields which we will
designate as $m_i$ and $\mu_{in}$ ($i<n$). This integral should be taken with
the weight $\exp({\rm i}\int {\rm d}t\, {\cal L})$ where the Lagrangian is
\begin{eqnarray} &&
  {\cal L}=\sum\limits_{a=1}^{d}m_a
    \left[\partial_t\rho_a+D\frac{d(d-2a+1)}2 \right]
    +\frac{{\rm i}D}{2}\left[d\sum\limits_a m_a^2
    -\left(\sum\limits_a m_a\right)^2\right]
 \nonumber \\ &&
  +{\rm i}Dd \sum\limits_{i<j}\exp(2\rho_j-2\rho_i)\mu_{ij}^2
    +2{\rm i}Dd\sum\limits_{i<k<j}\mu_{ij}\mu_{ik}
    \exp(2\rho_k-2\rho_i)\eta_{kj}
 \nonumber \\ &&
    +\sum\limits_{i<j}\mu_{ij}\partial_t\eta_{ij}
    +{\rm i}Dd\sum\limits_{i<k<m,n}
    \mu_{im}\mu_{in}\eta_{km}\eta_{kn}
    \exp(2\rho_k-2\rho_i) \,.
\label{gy8} \end{eqnarray}
Since the matrix $\hat B$ in accordance with (\ref{gy1}) does not
depend on $\hat R$ it is enough to know the statistics of $\rho_a$
and $\eta_{ij}$ to determine the average (\ref{gaa5}). Therefore,
instead of (\ref{ga16}) we get
\begin{eqnarray}
  && {\cal Z}(y)=\int {\cal D}\rho{\cal D}\eta{\cal D}m{\cal D}\mu\,
    \exp\left[\int_{-\infty}^0 \!\! {\rm d}t\, \
    \left( {\rm i}{\cal L}-\frac{y^2}{2}U\right)\right] \,.
  \label{gaa16}
\end{eqnarray}
Here $U$ is determined by (\ref{ga18}) where (\ref{gy1},\ref{gy5})
should be substituted.

Thus we obtained the expression for the generating functional
(\ref{gaa3}) in terms of the functional (path) integral which is
convenient for an analysis which is presented in the subsequent section.

\section{Generating Functional}
\label{sec:fun}

Here, we are going to calculate the generating functional (\ref{gaa3}) for a
single-point statistics of $\theta$ that is of the object (\ref{theta})
corresponding to $\beta({\bbox r})=\delta_\Lambda ({\bbox r})$  and also the
statistics of the difference that is of the object (\ref{diff}) corresponding
to (\ref{difb}). The starting point for the subsequent consideration is the
expression (\ref{gaa16}). There are different ways to examine ${\cal Z}(y)$.
We will describe two schemes leading to the same answer but carrying in some
sense a complimentary information. We believe also that the consideration of
the different schemes is useful from the methodical point of view. A
modification of the second scheme is presented in Appendix.

\subsection{Saddle-Point Approach}

The first way to obtain the answer for the generating functional
(\ref{gaa3}) is in using the saddle-point approximation for the path
integral (\ref{gaa16}). The inequalities justifying the
approximation are $\Lambda L\gg1$ for the object (\ref{theta}) and
$\Lambda r\gg1$ for the object (\ref{diff}).

As we will see large values of the differences $\rho_i-\rho_k$ ($i<k$) will
be relevant for us. At the condition fluctuations of $\eta$ and $\mu$ are
suppressed and it is possible to neglect the fluctuations. Therefore we can
omit the integration over $\eta$ and $\mu$ in (\ref{gaa16}) substituting
$\eta=\mu=0$ into (\ref{gy8}). After that we obtain a reduced Lagrangian.
Introducing an auxiliary field $s$ we can rewrite the reduced Lagrangian as
\begin{equation}
  {\rm i}{\cal L}_{\rm r} = {\rm i} \sum_{a=1}^{d} m_a
    \left[\partial_t\rho_a + D\frac{d(d+1-2a)}2+s\right]
    -\frac{Dd}{2}\sum\limits_a m_a^2-\frac{s^2}{2D} \,.
  \label{gy10}
\end{equation}
Now, to obtain ${\cal Z}(y)$ one should integrate the exponent in
(\ref{gaa16}) (with ${\cal L}_{\rm r}$) over $\rho_a$, $m_a$ and $s$. To
examine (\ref{gy10}) it is convenient to pass to new variables
$\phi_a=O_{ab}\rho_b$ and $\tilde m_a=O_{ab}m_b$ where $\hat O$ is an
orthogonal matrix. We make the following transformation
\begin{eqnarray} &&
  \phi_1=\sqrt{\frac{3}{d(d^2-1)}}\,
    \left[(d-1)\rho_1+(d-3)\rho_2+\dots +(1-d)\rho_d\right] \,,
\label{phi} \\ &&
 \phi_2= \dots \,, \quad  \dots \,, \quad
  \phi_d=\frac1{\sqrt{d}}\,
   \left[\rho_1+\rho_2+\dots +\rho_d\right] \,.
\nonumber \end{eqnarray}
Then the expression (\ref{gy10}) will be rewritten as
\begin{equation}
  {\rm i}{\cal L}_{\rm r} = {\rm i} \sum_{a=1}^{d}\tilde
    m_a\partial_t\phi_a +{\rm i}\sqrt{d}\tilde m_d s-\frac{s^2}{2D}
     -\frac{Dd}{2}\sum\limits_a \tilde m_a^2
     +{\rm i}\frac{Dd}{2}\sqrt{\frac{d(d^2-1)}{3}}\tilde m_1 \,.
\label{gy13} \end{equation}

The Lagrangian (\ref{gy13}) is a sum over different degrees of freedom. The
dynamics of $\phi_1$ is ballistic whereas the dynamics of $\phi_a$ for $a>1$
is purely diffusive. We will see that times determining the main contribution 
to the generating functional are large enough so that for the relevant region 
$\phi_1\gg\phi_a$. Therefore the potential $U$ (\ref{ga18}) depends
practically only on $\phi_1$ and it is possible to integrate explicitly over
$s$, $\phi_a$ and $\tilde m_a$ for $a>1$. After that we stay only with one
degree of freedom which is described by the Lagrangian
\begin{equation}
  {\rm i}{\cal L}_{1}= {\rm i}\tilde m_1\left(\partial_t\phi_1
    +\frac{Dd}{2}\sqrt{\frac{d(d^2-1)}{3}}\right)
    -\frac{Dd}{2}\tilde m_1^2 \,.
  \label{gy14}
\end{equation}

Neglecting all $\phi_a$ for $a>1$ and performing the transformation
inverse to (\ref{phi}) we obtain
\begin{eqnarray} &&
  \rho_1\approx\sqrt{\frac{3(d-1)}{d(d+1)}}\,\phi_1 \,, \qquad
   \rho_a\approx\frac{d-2a+1}{d-1}\rho_1 \,.
\label{conn} \end{eqnarray}
We will soon see that the characteristic value $\phi_1\gg1$. Therefore the 
characteristic value of $e^{\rho_1}$ is much larger than other $e^{\rho_a}$ 
and we conclude that the potential $U$ depends really only on $\rho_1$. For 
the case of the single-point statistics the characteristic value of the 
difference ${\bbox r}_1-{\bbox r}_2$ in (\ref{ga18}) is $\Lambda^{-1}$. Then 
it follows from (\ref{gy1},\ref{gy5}) that the potential $U$ falls down from 
$P_2$ to zero near the point $\rho_1=\ln(L\Lambda)$ that is near the point 
$\phi_1=\phi_\Lambda$ where
\begin{equation}
  \phi_\Lambda= \sqrt
    {\frac{d(d+1)}{3(d-1)}}{\rm ln\,}(L\Lambda) \,.
\label{jump} \end{equation}
For the difference the potential increases from zero to $2P_2$ at
$\phi_1=\phi_R$ where
\begin{equation}
  \phi_R= \sqrt
    {\frac{d(d+1)}{3(d-1)}}{\rm ln\,}(L/r_0) \,,
\label{jump1} \end{equation}
and then falls down from $P_2$ to zero near the point $\phi_1=\phi_\Lambda$.
The expressions (\ref{jump},\ref{jump1}) determine the characteristic values
of $\phi_1$ which are really large since $L\Lambda\gg1$ or $L/r_0\gg1$ what
justifies our conclusions.

Now we can employ the saddle-point approximation:
\begin{equation} \left.
  \ln{\cal Z}(y)\approx \int_{-\infty}^0 \!\! {\rm d}t\, \
    \left( {\rm i}{\cal L}_1-\frac{y^2}{2}U\right)\right|_{\rm inst} \,,
\label{go} \end{equation}
where we should substitute solutions of the extrema conditions which
we will call instantonic equations. The instantonic equations which can be
found as extrema conditions for ${\rm i}{\cal L}_1-y^2 U/2$ are
\begin{eqnarray} &&
  \partial_t\phi_1+\frac{Dd}{2}\sqrt{\frac{d(d^2-1)}{3}}
    =-{\rm i}Dd\tilde m_1 \,,
\label{go1} \\ &&
  \partial_t\tilde m_1
    ={\rm i}\frac{y^2}{2}\frac{\partial U}{\partial\phi_1} \,.
\label{go2} \end{eqnarray}
The equations conserve the `energy'
\begin{eqnarray} &&
  -{\rm i}\frac{Dd}{2}\tilde m_1\sqrt{\frac{d(d^2-1)}{3}}
    +\frac{Dd}{2}\tilde m_1^2+\frac{y^2}{2}U \,.
  \label{go3}
\end{eqnarray}
The conservation law is satisfied since ${\rm i}{\cal L}_1-y^2U/2$
does not explicitly depend on $t$. The `energy' (\ref{go3}) is
equal to zero since at $t\to-\infty$ the value of $\tilde m_1$ should
tend to zero. The property can be treated as the extremum condition
appearing at variation of ${\rm i}{\cal L}_{\rm r}-y^2 U/2$ over the
initial value of $\phi_1$. Equating (\ref{go3}) to zero we can
express $\tilde m_1$ via $\phi_1$. Next, since (\ref{go3}) is zero
then the saddle-point value of ${\cal Z}(y)$ (\ref{go}) can be
written as ${\rm i}\int{\rm d}\phi_1\,\tilde m_1$, where the integral
over $\phi_1$ goes from zero to infinity.

Substituting into ${\rm i}\int{\rm d}\phi_1\,\tilde m_1$ the expression of
$\tilde m_1$ in terms of $\phi_1$ we get for the single-point statistics
\begin{equation}
  \ln{\cal Z}(y)\simeq \frac{d(d+1)}{6}
    \left[1-\sqrt{1+\frac{12y^2 P_2}{Dd^2(d^2-1)}}\,\right]
    {\ln}(L\Lambda) \,.
  \label{gy15}
\end{equation}
Note that the expression (\ref{gy15}) has (as a function of $y$) two
branch points $y=\pm{\rm i}y_{\rm sing}$ where
\begin{equation}
  y_{\rm sing}^2 = \frac{Dd^2(d^2-1)}{12 P_2} \,.
  \label{sing}
\end{equation}
The same procedure can be done for the passive scalar difference, or,
more precisely, for the object (\ref{diff}). Taking into account the presence
of the jumps (\ref{jump1}) and (\ref{jump}) in the potential $U$ we get an
answer slightly different from (\ref{gy15})
\begin{eqnarray} &&
  \ln{\cal Z}(y)\simeq \frac{d(d+1)}6
    \left[1-\sqrt{1+\frac{24y^2 P_2}{Dd^2(d^2-1)}}\,\right]
    {\ln}(r_0\Lambda) \,,
 \label{go4} \\ &&
   y_{\rm sing}^2 = \frac{Dd^2(d^2-1)}{24 P_2} \,.
\label{goo4} \end{eqnarray}
Note that (\ref{go4}) does not depend on the pumping scale $L$ but
still depend on the cutoff $\Lambda$.

The characteristic value of $\phi_1$ is determined by the quantity
(\ref{jump}) which is much larger than unity. Then it follows from
(\ref{conn}) that $\exp(2\rho_j-2\rho_i)\ll1$, $i>j$, (excluding for a short 
initial stage of the evolution) and we see from (\ref{gy8}) that fluctuations 
of the fields $\eta$ are suppressed in comparison, say, with $\rho_a$. This 
justifies neglecting the fields $\eta$ and $\mu$ leading to the
reduced Lagrangian (\ref{gy10}). Next, dynamics of $\phi_a$ for $a>1$ is
diffusive and it follows from (\ref{gy13}) that the characteristic value of
$\phi_a$ can be estimated as $\sqrt{Dd|t|}$. As follows from (\ref{gy13})
$\partial_t\phi_1\sim Dd^{5/2}$ and we find from (\ref{jump}) the instantonic
life time
\begin{equation}
  t_{\rm lt}=D^{-1}d^{-2}\ln(L\Lambda) \,,
 \label{life} \end{equation}
which determines times producing nonzero contribution to the effective
action. At $|t|\sim t_{\rm lt}$ the characteristic values of $\phi_a$ for
$a>1$ are of order $\sqrt{\ln(L\Lambda)/d}$ and we conclude that
 \begin{eqnarray} &&
  \frac{\phi_a}{\phi_1}\sim \frac{1}{d\sqrt{\ln(L\Lambda)}}\ll1 \,,
 \label{goj} \end{eqnarray}
at times $|t|\sim t_{\rm lt}$. The inequality (\ref{goj}) justifies passing
to the Lagrangian (\ref{gy14}). The same arguments can be applied to the
generating functional for the passive scalar difference, the only
modification is in the substitution $\ln(L\Lambda)\to{\ln}(r_0\Lambda)$.

There are also additional applicability conditions for the answers
(\ref{gy15},\ref{go4}). To establish the conditions one should go beyond the
main order of the saddle-point approximation. It will be more convenient for
us to develop an alternative scheme which enables to find the conditions
simpler. That is the subject of the next subsection.

\subsection{Schr\"odinger equation}

Here we present another way to get the answers (\ref{gy15},\ref{go4}).
As previously we start from the path integral representation (\ref{gaa16})
for the generation functional ${\cal Z}(y)$.

Unfortunately it is impossible to get a closed equation for ${\cal Z}(y)$.
To avoid the difficulty we introduce an auxiliary quantity
\begin{equation}
  \Psi(t,y,\rho_0,\eta_0)=\left.\int
    {\cal D}\rho{\cal D}\eta{\cal D}m{\cal D}\mu\,
    \exp\left[\int_{-t}^0 \!\! {\rm d}t' \,
    \left( {\rm i}{\cal L}-\frac{y^2}{2}U\right)\right]
    \right|_{\rho(-t)=\rho_0,\eta(-t)=\eta_0} \,.
  \label{ga22}
\end{equation}
It follows from the definition (\ref{ga22}) that
\begin{equation}
  {\cal Z}(y)=\lim_{t\to\infty}\int\prod {\rm d}\rho_a
    \, {\rm d}\eta_{ij}\,
    \Psi(t,y,\rho,\eta) \,.
  \label{ga23}
\end{equation}
The equation for the function $\Psi$ can be obtained from the
expression (\ref{gy8}) and the definition (\ref{ga22}):
\begin{eqnarray}
  && \partial_t \Psi = \frac{Dd}2 \left[
    \sum\limits_{i=1}^{d}\frac{\partial^2}{\partial\rho_i^2}
    -\frac{1}d\left(\sum\limits_{i=1}^{d}
    \frac{\partial}{\partial\rho_i}\right)^2
    -\sum\limits_{i=1}^{d}(d-2i+1)\frac{\partial}{\partial\rho_i}
    \right. \nonumber \\
  && + 2\sum\limits_{i<j}\exp(2\rho_j-2\rho_i)
    \frac{\partial^2}{\partial\eta_{ij}^2}
    +4\sum\limits_{i<k<j}
    \exp(2\rho_k-2\rho_i)
    \frac{\partial}{\partial\eta_{ij}}
    \frac{\partial}{\partial\eta_{ik}}\eta_{kj}
    \nonumber \\
  && \left. + 2\sum\limits_{i<k<m,n}
    \exp(2\rho_k-2\rho_i) \frac{\partial}{\partial\eta_{im}}
    \frac{\partial}{\partial\eta_{in}}\eta_{km}\eta_{kn}\right] \Psi
    -\frac{y^2 U}2 \Psi \,.
  \label{gyy9}
\end{eqnarray}
We see that the equation (\ref{gyy9}) for $\Psi$ resembles the
Schr\"odinger equation. The initial condition to the equation can be found
directly from the definition (\ref{ga22}):
\begin{equation}
  \Psi(t=0,y,\rho,\eta)=\prod\delta(\rho_a)\delta(\eta_{ij}) \,.
  \label{gaa22}
\end{equation}
The value of ${\cal Z}$ in accordance with (\ref{ga23}) is determined by the
integral of $\Psi$ over $\eta$ and $\rho$. This integral is equal to unity at
$t=0$ and then varies with increasing time $t$ due to $U\neq0$ since only the
term with $U$ in (\ref{gaa22}) brakes conservation of the integral. Thus to
find ${\cal Z}$ we should establish an evolution of the function $\Psi$ from
$t=0$ to large $t$.

Below we concentrate on the single-point statistics. The scheme can be
obviously generalized for the passive scalar difference.

Let us first describe the evolution qualitatively.
The initial condition (\ref{gaa22}) shows that at $t=0$ the function $\Psi$
is concentrated at origin. Then it undergoes spreading in all directions
except for $\rho_1+\dots+\rho_d$ since the operator in the right-hand side of
(\ref{gyy9}) commutes with $\rho_1+\dots+\rho_d$. This is a consequence of the
condition ${\rm Det}\,\hat T=1$ (to be satisfied) which implies that during
the evolution $\rho_1+\dots+\rho_d=0$. That means that we should treat a
solution of (\ref{gyy9}) which is $\Psi\propto\delta(\rho_1+\dots+\rho_d)$.
The function $\Psi$ is smeared diffusively with time and also moves as a
whole in some direction, which is determined by the term with the first
derivative in (\ref{gyy9}). The rate of the ballistic motion is
\begin{equation}
  \langle \partial_t \rho_i \rangle = D\,\frac{d(d-2i+1)}2 \,.
  \label{psi1}
\end{equation}
Therefore $\Psi$ describes a cloud, centre of which moves according to
the law
\begin{equation}
  \rho_i = D\,\frac{d(d-2i+1)}2 t\,.
  \label{ps1}
\end{equation}
Effective diffusion coefficients for $\eta$'s fall down with increasing $t$
since in accordance with (\ref{ps1}) the differences $\rho_k-\rho_i$
[figuring in (\ref{gyy9})] are negative and grow by their absolute
value. Therefore the diffusion over $\eta$ stops when the
characteristic values of $\rho_i-\rho_k$ becomes greater than unity.
Note that the `frozen' values of $\eta$ do not depend on $y$ since
$U$ can be considered as uniform during the initial stage of the
evolution. After that $\eta$ are frozen, the diffusion continues only
over $\rho$'s. If the cloud is inside the region where $U \simeq P_2$ then
the evolution of the cloud is not influenced by $U$. After the period of
time $t_{\rm lt}$ (\ref{life}) the cloud reaches a barrier where the
potential $U$ falls down from $P_2$ to $0$. The subsequent history depends on
the value of $y$. For moderate $y$ the cloud passes this barrier and
continues to move with the same rate. After this the integral of $\Psi$ will
not change in time, and its value will determine the generating functional
${\cal Z}(y)$.  Naive estimations would give the answer
$\ln{\cal Z}(y)=-y^2 t_{\rm lt}/2$, which reproduces the value of the pair
correlation function of $\theta$.

A special consideration is needed if $|y|\gg y_{\rm sing}$ or if $y$ is close
to $\pm {\rm i}y_{\rm sing}$ where $y_{\rm sing}$ is introduced by
(\ref{sing}). Just the last region determined the PDF's and is consequently
of a special interest. Note that $y=\pm {\rm i}y_{\rm sing}$ corresponds to
appearance of a bound state near the pumping boundary (where $U$ falls down
from $P_2$ to zero). If $y\gg y_{\rm sing}$ then the front side of the cloud
reaches the jump of the potential much earlier than $t_\Lambda$. The residue
of the cloud (living inside the potential well) is damped due to the term
with $y$ and don't give a contribution to ${\cal Z}(y)$. If
$|y|\gg y_{\rm sing}$ then ${\cal Z}(y)\gg \exp(-y^2 t_\Lambda/2)$, really
the asymptotics of ${\cal Z}(y)$ is exponential in the case. If
$|y \pm {\rm i}y_{\rm sing}| \ll y_{\rm sing}$ then the cloud stays near the
pumping boundary for a long time, that is the shape of $\Psi$ inside the
region $U\simeq P_2$ varies in time comparatively slow. Besides, a
part of $\Psi$ percolates out to the region where $U\simeq 0$ and the
integral of $\Psi$ grows with increasing $|t|$. As $y$ come to
${\rm i}y_{\rm sing}$ closer this stage lasts longer. One can say that the
back side of the cloud $\Psi$ gives the right answer for ${\cal Z}(y)$. The
important point is that if $y$ is not very close to ${\rm i}y_{\rm sing}$
then during the time of exiting $\Psi$ from the potential the width of $\Psi$
in terms of diffusive degrees of freedom is much less than
${\rm ln\,}L\Lambda$. This means that the function $\Psi$ is really narrow,
that justifies our consideration.

For an quantitative analysis it is convenient to pass to the
variables $\phi_i$ (\ref{phi}). Since an $\eta$-dependence of $\Psi$ is frozen
after an initial part of evolution, then it is possible to obtain an
equation for the integral of $\Psi$ over $\eta$:
\begin{equation}
  \tilde\Psi(\phi_1, \dots, \phi_{d-1}) = \int{\rm d}\phi_d \prod
    {\rm d}\eta_{ij}\,\Psi \,,
  \label{int}
\end{equation}
where we included also an integration over $\phi_d$ to remove the
factor $\delta(\rho_1+\dots+\rho_d)$. The equation for the function
(\ref{int}) is
\begin{equation}
  \partial_t \tilde\Psi = \frac{Dd}2 \left[ \sum_{i=1}^{d-1}
    \frac{\partial^2}{\partial\phi_i^2} - \sqrt{\frac{d(d^2-1)}3}
    \frac{\partial}{\partial \phi_1} \right] \tilde\Psi - \frac{y^2
    \tilde U}2 \tilde\Psi \,,
  \label{dpsi}
\end{equation}
where $\tilde U$ is function of $\phi_a$ only which can be found by
substituting into $U$ the `frozen' values of $\eta$'s. Qualitatively
$\tilde U$ has the same structure as $U$ itself. One can conclude from
(\ref{dpsi}) that the cloud described by $\tilde\Psi$ moves ballistically
into the $\phi_1$ direction and spreads along other directions. We are
going to treat the situation when the cloud remains narrow during
the relevant part of the evolution. Then one can integrate $\tilde\Psi$ over
all $\phi_i$, $i>1$ in a similar way as in the case with $\eta$'s and get an
$1d$ equation for
\begin{eqnarray} &&
  \bar\Psi(\phi_1)= \int \prod\limits_{2}^{d-1} {\rm d}\phi_i\,
    \tilde\Psi \,.
\nonumber \end{eqnarray}
The function $\bar\Psi$ satisfies the equation
\begin{equation}
  \partial_t \bar\Psi = \frac{Dd}2 \left[
    \frac{\partial}{\partial\phi_1}
     - \sqrt{\frac{d(d^2-1)}3} \, \right]
     \frac{\partial}{\partial \phi_1} \bar\Psi -
    \frac{y^2 \bar U}2 \bar\Psi \,.
  \label{psi2}
\end{equation}
The initial condition to Eq. (\ref{psi2}) is $\bar\Psi(t=0)=\delta(\phi_1)$.
The potential $\bar U$ is obtained from $\tilde U$ by substitution
$\phi_a\to0$ for $a>0$. In fact, on the direction (\ref{ps1}) we have that is
the potential $\bar U$ depends only on $\rho_1$. The barrier is achieved when
$\rho_1 \simeq {\rm ln\,}L\Lambda$. Passing to the variables $\phi_i$ we
conclude that the potential $\bar U$ diminishes from $P_2$ at
$\phi_1<\phi_\Lambda$ to zero at $\phi_1>\phi_\Lambda$ where $\phi_\Lambda$
is defined by (\ref{jump}).

The character of a solution of the equation (\ref{psi2}) can be analyzed
semiqualitatively in terms of the width $l$ of $\bar\Psi$ over $\phi_1$
and its amplitude $h$. When $\bar\Psi$ reaches the pumping boundary then it
stops there for a period of time. Then the width $l$ and the amplitude $h$
governed by the equations
\begin{equation}
  \frac{{\rm d}l}{{\rm d}t} = - Dd\lambda + \frac{Dd}{l}
    \,, \quad \frac{{\rm d}h}{{\rm d}t} = - \frac{Dd
    h}{l^2} - \frac{y^2 P_2 h}2 \,,
\end{equation}
where $\lambda = \sqrt{d(d^2-1)/12}$, $Dd\lambda$ is the
rate of the cloud motion along $\phi_1$ direction (when
$U={\rm const}$), and $Dd$ is the diffusion coefficient for
$\phi_1$ direction. One can estimate from the 1st equation
the width $l$ as $l \sim 1/\lambda$. Then from the 2nd
equation the height $h$ falls down or grows in time
depending on $y$. The characteristic $y$ where the regime
changes is of the order $|y_{\rm sing}|^2 \sim
Dd\lambda^2/P_2$. We'll show this by consistent
calculations.

The equation (\ref{psi2}) can be solved analytically, e.g., by the Laplace
transform over time $t$. Making the Laplace transform one gets
\begin{equation}
  p \bar\Psi(p) - \delta(\phi_1) = \frac{Dd}2 \left[
    \frac{\partial}{\partial\phi_1} -
    \sqrt{\frac{d(d^2-1)}3} \, \right]
    \frac{\partial}{\partial \phi_1} \bar\Psi(p) -
    \frac{y^2}{2}\bar U(\phi_1) \bar\Psi(p) \,.
  \label{psi2p}
\end{equation}
We are interesting in the bound state describing by the equation.
Solutions for $\bar\Psi(p)$ in the intervals $(-\infty, 0)$,
$(0,\phi_\Lambda)$, $(\phi_\Lambda, \infty)$ are exponential, one should
match them. The function $\bar\Psi(p)$ as a function of $p$ has two
branch points at
\begin{eqnarray} &&
  p_1 = -\frac{Dd^2(d^2-1)}{24} - y^2 P_2/2 \,, \qquad
  p_2= -\frac{Dd^2(d^2-1)}{24} \,, \label{bpp}
\end{eqnarray}
coming from the regions $\phi_1 < \phi_\Lambda$ and $\phi_1>\phi_\Lambda$,
respectively. When one of this branch points passes $p = 0$ then $\bar\Psi$
starts to grow exponentially in time. This happens when $y$ passes
$\pm{\rm i}y_{\rm sing}$ from (\ref{sing}) moving along the imaginary axis.

The value of the generating functional is determined in accordance
with (\ref{ga23}) by large time behavior of $\Psi(t)$. This means
that we should be interested in the behavior of $\Psi(p)$ at small
$p$. Really the function $\int{\rm d}\phi_1\,\hat\Psi(p)$ entering
(\ref{ga23}) has a pole at $p=0$ related to the asymptotic behavior
\begin{eqnarray} &&
  \bar\Psi(p)\propto\exp\left(-\frac{2p}{Dd}
   \sqrt{\frac{3}{d(d^2-1)}}\phi_1\right) \,,
\nonumber \end{eqnarray}
at $\phi_1>\phi_\Lambda$ and small $p$, the behavior can be found from
(\ref{psi2p}).  Just the residue of $\int{\rm d}\phi_1\,\bar\Psi(p)$
at the pole determines ${\cal Z}(y)$. To find the value of the
residue we should analyze the behavior of $\bar\Psi(p)$ at
$0<\phi_1<\phi_\Lambda$. At small $p$ there are two contributions into
$\bar\Psi$ which behave as
\begin{eqnarray} &&
  \propto\exp \left\{ \left( \sqrt{\frac{d(d^2-1)}{12}}
     \pm\sqrt{\frac{d(d^2-1)}{12}+\frac{y^2P_2}{Dd}}\right)
   \phi_1\right\} \,,
\label{inter} \end{eqnarray}
as follows from (\ref{psi2p}) at $p=0$. Therefore the residue which
is determined by the integral $\int{\rm d}\phi_1\,\bar\Psi(p)$ over
the region $\phi_1>\phi_\Lambda$ is proportional to
\begin{eqnarray} &&
  \exp\left\{\left(\sqrt{\frac{d(d^2-1)}{12}}+
   \sqrt{\frac{d(d^2-1)}{12}
    +\frac{y^2P_2}{Dd}}\right)\phi_\Lambda\right\} \,,
\label{go8} \end{eqnarray}
Substituting here (\ref{jump}) we reproduce (\ref{gy15}).

Let us now establish the applicability condition of the above procedure.
The expression (\ref{go8}) implies that the exponent with the sign minus in
(\ref{inter}) gives a negligible contribution to $\Psi(p)$ at
$\phi_1=\phi_\Lambda$. The condition is satisfied if
\begin{eqnarray} &&
  |y^2+y_{\rm sing}^2|\phi_\Lambda^2\gg
   \frac{Dd}{P_2} \,.
\nonumber \end{eqnarray}
Substituting here (\ref{jump},\ref{sing}) we obtain
\begin{equation}
  \left| \frac{y \pm {\rm i}y_{\rm sing}}{y_{\rm sing}}\right| \gg
   \left( d^4\,{\ln}^2 L\Lambda \right)^{-1} \,.
\label{go9} \end{equation}
For $y$ closer to $\pm {\rm i}y_{\rm sing}$ one should be careful since then
a fine analytical structure of ${\cal Z}(y)$ will be relevant. As an analysis
for $d=2$ shows \cite{98Ste} ${\cal Z}(y)$ has a system of poles along the
imaginary semiaxis starting from $\pm {\rm i}y_{\rm sing}$ and the parameter
$\left( d^4{\ln}^2 L\Lambda \right)^{-1}$ just determines the separation
between the poles. The assertion about the cut made in the previous
subsection is related to the restrictions of the saddle-point approximation
which cannot feel this fine pole structure, it gives a picture averaged over
the interpole distances, it is precisely the cut. This approach is correct
just at the condition (\ref{go9}).

Note that the same criterion (\ref{go9}) justifies our assumption that the
cloud described by $\Psi$ is narrow during the relevant part of the
evolution. Namely, the duration of the part is determined by the time
$t_{\rm exit}=p_1^{-1}$ (see (\ref{bpp})). This is the time which the cloud
stays near the barrier. For $y$ close to $\pm iy_{\rm sing}$ the time can be
estimated as $t_{\rm exit}^{-1}\sim P_2|y_{\rm sing}| |y\mp iy_{\rm sing}|$.
Then the diffusive width $\sqrt{Dd t_{\rm exit}}$ of $\Psi$ in the directions
$\phi_a$ for $a>1$ is much less than $\phi_\Lambda$ just if (\ref{go9}) is
satisfied. Principally the diffusive dynamics at $d>2$ could modify the
noted fine pole structure of ${\cal Z}$, the problem needs a separate
investigation.

The same procedure can be done for the passive scalar differences. The
cloud $\Psi$ should pass the region $\rho_1<\ln(L/r_0)$ before it reaches the
potential. Then it enters the region $\bar U =2P_2$ with some finite
diffusive width. One can note, however, that it is irrelevant. The only
characteristics of potential which are needed are its value (here $2P_2$
instead of $P_2$) and the length of the path inside it (which is
$\Delta\rho_1 = \ln(r_0\Lambda)$ instead of $\ln(L\Lambda)$).The evolution of
$\bar\Psi$ goes in the same way as in the case of single-point statistics.
Again, we get the answer (\ref{go4}) and the criterion analogous to
(\ref{go9}).

In the subsection we presented the analysis based on the dynamical equation
(\ref{gyy9}) for the auxiliary object $\Psi$. The results obtained can be
reproduced also on an alternative language: For this we should introduce
another auxiliary object the equation for which is stationary. The
corresponding scheme which could be interesting from the methodical point of
view is sketched in Appendix.

\section{Calculation of PDF}
\label{sec:pdf}

In this section we calculate the PDF's ${\cal P}$ for the objects
(\ref{theta}) and (\ref{diff}). The most convenient way to do it is
in using the relation
\begin{equation}
  {\cal P}(\vartheta)=\int\frac{{\rm d}y}{2\pi}
    \exp(-{\rm i}y\vartheta) {\cal Z}(y) \,,
  \label{gy16}
\end{equation}
where $\vartheta$  is
\begin{equation}
  \vartheta=\int {\rm d}{\bbox r}\, \beta({\bbox r})
    \theta(0,{\bbox r}) \,.
  \label{ga4}
\end{equation}
Let us remind that knowing ${\cal P}(\vartheta)$ one can restore also
moments of $\vartheta$:
\begin{eqnarray} &&
  \langle|\vartheta|^n\rangle
   =\int{\rm d}\vartheta\,|\vartheta|^n{\cal P}(\vartheta) \,.
\label{mom} \end{eqnarray}

The generating functional in (\ref{gy16}) is determined by (\ref{gy15}) or
(\ref{go4}). Being interested in the main exponential dependence of the PDF's
for the objects (\ref{theta}) and (\ref{diff}) we can forget about
preexponents. Then
\begin{eqnarray} &&
 {\cal P}(\vartheta)=
  \int \frac{{\rm d}y}{2\pi} {\rm exp} \left(
    - {\rm i}y\vartheta + q \left[ 1 - \sqrt{1+
    y^2/y_{\rm sing}^2 } \right] \right) \,,
\label{ygy16} \end{eqnarray}
where
\begin{eqnarray} &&
  y_{\rm sing}^2 = \frac{Dd^2(d^2-1)}{12 P_2}\ \ ({\rm
    for\ scalar}), \quad y_{\rm sing}^2 =
    \frac{Dd^2(d^2-1)}{24 P_2}\ \ ({\rm for\ difference}),
    \\ &&
     q=\frac{d(d+1)}6 \, \ln(L\Lambda)\ \ ({\rm for\
    scalar}), \quad q=\frac{d(d+1)}6 \, \ln(r_0\Lambda)\ \
    ({\rm for\ difference}). \label{qu}
\end{eqnarray}
Since both $q$ defined by (\ref{qu}) are regarded to be much larger than
unity the integral (\ref{ygy16}) can be calculated in the saddle-point
approximation. The saddle-point value is
\begin{eqnarray} &&
  y_{\rm sp} = {\rm i}\frac{y_{\rm sing}}
   {1+q^2/y_{\rm sing}^2 \vartheta^2} \,.
\label{sadd} \end{eqnarray}
Then
\begin{equation}
  {\rm ln\,} {\cal P}(\vartheta) \simeq q \left( 1 -
    \sqrt{1+\frac{y_{\rm sing}^2 \vartheta^2}{q^2}} \right) \,.
\label{lnpdf} \end{equation}
The expression leads to the exponential tail
\begin{eqnarray} &&
 {\rm ln\,} {\cal P}(\vartheta)
  \simeq -y_{\rm sing}|\vartheta| \,,
\label{tail} \end{eqnarray}
realized at $|\vartheta|\gg q/y_{\rm sing}$. The coefficient $y_{\rm
sing}$ in (\ref{tail}) determined by (\ref{sing}) is in accordance
with the result obtained in \cite{98BGK}.

The expression (\ref{lnpdf}) enables one to find the following
averages in accordance with (\ref{mom})
\begin{eqnarray} &&
 \langle\theta_\Lambda^2\rangle=
  \frac{2P_2}{d(d-1)D}\ln(L\Lambda) \,, \quad
   \langle(\Delta\theta_\Lambda)^2\rangle=
    \frac{4P_2}{d(d-1)D}\ln(r_0\Lambda) \,.
\label{pair} \end{eqnarray}
The expressions (\ref{pair}) can be obtained also by direct expansion
of ${\cal Z}(y)$ from (\ref{gy15}) or (\ref{go4}). The universal tail
(\ref{tail}) is realized if
\begin{eqnarray} &&
 \theta_\Lambda\gg \sqrt{\langle\theta_\Lambda^2\rangle}\,
   d\ln(L\Lambda) \,, \quad
 \Delta\theta_\Lambda\gg \sqrt{\langle(\Delta\theta_\Lambda)^2\rangle}\,
   d\ln(r_0\Lambda) \,.
\label{crit} \end{eqnarray}
Since both logarithms are supposed to be large we conclude that there
exists a relatively wide region where the statistics of $\vartheta$
is approximately Gaussian, the region is determined by the
inequalities inverse to (\ref{crit}).

Let us discuss the applicability conditions of the expression (\ref{lnpdf}).
First, if one calculates the passive scalar PDF by the saddle point method
then the position of the saddle point is determined by (\ref{go9}) if
\begin{equation}
  \vartheta \ll d^2 \sqrt{\frac{P_2}D} \,
    {\ln}^2 (L\Lambda) \,.
 \label{remo} \end{equation}
The applicability domain of the saddle-point method overlaps the
region of validity of expression (\ref{gy15}) for the generation 
function ${\cal Z}(y)$.  The above inequalities are correct for 
$\theta_\Lambda$, for $\Delta\theta_\Lambda$ one should substitute 
$\ln(L\Lambda)$ by $\ln(r_0\Lambda)$. Second, fluctuations of $y$ have to be 
small compared to the distance between $y_{\rm sp}$ and $y_{\rm sing}$. This 
gives the same criterion (\ref{remo}).

Let us stress that though formally our procedure is incorrect at
$\vartheta \gtrsim d^2\sqrt{P_2/D} \,{\ln}^2 (L\Lambda)$
the answer will be the same: the PDF will be determined by the exponential
tail (\ref{tail}). The point is that the character of the integral
(\ref{gy16}) at such extremely large $\vartheta$ will be determined by the
position of the singular point of ${\cal Z}(y)$ nearest to the real axis.
This is just ${\rm i}y_{\rm sing}$ leading to (\ref{tail}). To conclude, only
the character of the preexponent in ${\cal P}(\vartheta)$ is changed at
$\vartheta \sim d^2 \sqrt{P_2/D} \,{\ln}^2 (L\Lambda)$ whereas the
principal exponential behavior of ${\cal P}(\vartheta)$ remains unchanged
there.

\section*{Conclusion}

The single-point statistics of the passive scalar $\theta$ and the statistics
of its difference $\Delta\theta$ are traditional objects which carry an
essential information about correlation functions of the passive scalar in
the convective interval.  We examined the passive scalar in the large-scale
turbulent flow where the correlation functions logarithmically depend on
scale. Since really the logarithms are not very large it is useful to have
the whole PDF's of $\theta$ and $\Delta\theta$. That was the main purpose of
our investigation which was performed in the frame of the Kraichnan model.
The single-point PDF for the passive scalar and the PDF for the passive
scalar differences can be obtained from (\ref{lnpdf}) if to substitute
$\Lambda\to r_{\rm dif}^{-1}$ where $r_{\rm dif}$ is the diffusive length.
Though both the advecting velocity and the pumping force in the Kraichnan
model are considered as $\delta$-correlated in time we hope that our answer
are universal that is are true in the limit when the size of the convective
interval tends to infinity for arbitrary temporal behavior of the velocity
and of the pumping. The reason is that the spectral transfer time grows with
increasing the convective interval and in the limit is much larger than the
correlation times of the velocity and of the pumping.

We believe also that the analytical scheme proposed in our work could be
extended for other problems related to the passive scalar statistics.
Note as an example the work \cite{98GK} where a modification of the scheme
enabled to find the statistics of the passive scalar dissipation. It is also
useful for investigating the large-scale statistics (on scales larger that
the pumping length) of the passive scalar \cite{98BFL}. We hope also that it
is possible to go beyond the case of the large-scale velocity field using the
perturbation technique of the type proposed in \cite{95SS,96PSS,97BFL}.

\acknowledgements

We are grateful to E.~Balkovsky, M.~Chertkov, G.~Falkovich, K.~Gawedzki and
M.~Olshanetsky for useful discussions. This work was supported in
part by Einstein and Minerva Centers at the Weizmann Institute, by
the grants of Minerva Foundation, Germany and Israel Science
Foundation, by Russian Foundation for Basic Research (I.K., M.S., gr.
98-02-17814), by Soros Foundation (M.S., gr. a98-674) and by INTAS
(M.S., gr. 96-0457) within the program of ICFPM.

\appendix

\section{}

Here we present an alternative way to obtain the answers
(\ref{gy15},\ref{go4}). We will use an auxiliary quantity
\begin{equation}
  \Xi(y,\rho_0,\eta_0)=\left.\int
    {\cal D}\rho{\cal D}\eta{\cal D}m{\cal D}\mu\,
    \exp\left[\int_{-\infty}^0 \!\! {\rm d}t \,
    \left( {\rm i}{\cal L}-\frac{y^2}{2}U\right)\right]
    \right|_{\rho(0)=\rho_0,\eta(0)=\eta_0} \,,
  \label{gu1}
\end{equation}
then
\begin{equation}
  {\cal Z}(y)=\Xi(y,0,0) \,.
  \label{gu2}
\end{equation}
Function $\Xi$ can be also defined as
\begin{equation}
  \Xi(y,\rho_0,\eta_0) = \lim_{t\to\infty}\int\prod {\rm
    d}\rho_a \, {\rm d}\eta_{ij}\, \Psi(t,y,\rho,\eta) \,,
  \label{gu2p5}
\end{equation}
where $\Psi$ is governed by equation (\ref{gyy9}) with
initial condition $\Psi(t=0,y,\rho,\eta)=\delta(\rho -
\rho_0)\delta(\eta - \eta_0).$
The equation for the function $\Xi$ can be found from
(\ref{gy8},\ref{gu1}):
\begin{eqnarray}
  &&  \left[ \sum\limits_{i=1}^{d}\frac{\partial^2}{\partial\rho_i^2}
    -\frac{1}d\left(\sum\limits_{i=1}^{d}
    \frac{\partial}{\partial\rho_i}\right)^2
    +\sum\limits_{i=1}^{d}(d-2i+1)\frac{\partial}{\partial\rho_i}
    \right. \nonumber \\
  && + 2\sum\limits_{i<j}\exp(2\rho_j-2\rho_i)
    \frac{\partial^2}{\partial\eta_{ij}^2}
    +4\sum\limits_{i<k<j}
    \exp(2\rho_k-2\rho_i)\eta_{kj}
    \frac{\partial}{\partial\eta_{ij}}
    \frac{\partial}{\partial\eta_{ik}}
    \nonumber \\
  && \left. + 2\sum\limits_{i<k<m,n}
    \exp(2\rho_k-2\rho_i)\eta_{km}\eta_{kn}
    \frac{\partial}{\partial\eta_{im}}
    \frac{\partial}{\partial\eta_{in}}\right] \Xi
    -\frac{y^2 U}{Dd} \Xi =0 \,.
  \label{gu3}
\end{eqnarray}
The boundary condition for the equation (\ref{gu3}) follows from the
definition (\ref{gu1}): For large enough $\rho_i,\,\eta_i$ the
potential $U=0$ for $t=0$ and remains zero also at finite times $t$.
Therefore the integral (\ref{gu1}) should be equal to unity in the
case. Thus $\Xi(y,\rho,\eta)$ should tend to unity where
$\rho,\,\eta\to\infty$.

Let us rewrite the equation (\ref{gu3}) in terms of the variables
(\ref{phi}):
\begin{eqnarray}
  && \big( \hat\Gamma_1 + \hat\gamma \big) (\Xi_1 + \xi) = 0 \,,
    \quad \Xi = \Xi_1 + \xi \,, \label{gu4} \\
  && \hat\Gamma_1 = \frac{\partial^2}{\partial\phi_1^2} +
    \sqrt{\frac{d(d^2-1)}3} \frac{\partial}{\partial\phi_1}
    - \frac{y^2 U}{Dd} \,, \\
  && \hat\gamma=\sum_{i=2}^{d-1} \frac{\partial^2}{\partial\phi_i^2}
    + 2\sum\limits_{i<k}\exp(2\rho_k-2\rho_i)
    \frac{\partial^2}{\partial\eta_{ik}^2} + 4\sum\limits_{i<k<n}
    \exp(2\rho_k-2\rho_i)\eta_{kn} \frac{\partial}{\partial\eta_{in}}
    \frac{\partial}{\partial\eta_{ik}} \nonumber \\
  && + 2\sum\limits_{i<k<m,n}
    \exp(2\rho_k-2\rho_i)\eta_{km}\eta_{kn}
    \frac{\partial}{\partial\eta_{im}}
    \frac{\partial}{\partial\eta_{in}}\,.
  \label{gu5}
\end{eqnarray}
Here $U$ as a function of $\phi_1$ is equal to $P_2$ inside a region
restricted by $\phi_\Lambda^-$ and $\phi_\Lambda^+$ (where $\phi_\Lambda^\pm$
are functions of variables $\phi_2, \dots, \phi_d, \eta$) and tends to zero
outside the region. We will solve the equation (\ref{gu4}) using the
perturbation theory over $\hat\gamma$, $\xi$. Then the zero order equation is
\begin{eqnarray} &&
  \hat\Gamma_1 \Xi_1 =0 \,.
\label{zero} \end{eqnarray}
The equation (\ref{zero}) can be easily solved at
$\phi_\Lambda^- < \phi_1 < \phi_\Lambda^+$, the answer is
\begin{equation}
  \Xi_1 \simeq \frac{2\lambda}{\sqrt{\lambda^2 + \frac{y^2
    P_2}{Dd}} + \lambda} {\rm exp} \left( - \left(
    \sqrt{\lambda^2 + \frac{y^2 P_2}{Dd}} - \lambda \right)
    (\phi_\Lambda^+ - \phi_1) \right) \,,
  \label{res}
\end{equation}
where $\lambda = \sqrt{d(d^2-1)/12}$, $Dd\lambda$ is the rate of the cloud
motion along $\phi_1$ direction. The result (\ref{res}) can be obtained using
the inequality $\sqrt{\lambda^2 + y^2 P_2/Dd} \, {\rm ln\,}L\Lambda \gg 1$.
The derivative $\partial \Xi_1 / \partial \phi_1 = 0$ at
$\phi_1<\phi_\Lambda^-$. However, $\Xi_1 \ne 1$ in this region. This is due
to the following fact: this region corresponds to the evolution of $\Psi$
when its initial position is to the left of potential $U$ (see
(\ref{gu2p5})). During the evolution cloud $\Psi$ passes the region of $U$
and its integral over $\rho$, $\eta$ changes. Then $\Xi$ is not equal to $1$.
Only when the distance between initial position and potential is of order
${\rm ln}^2 L\Lambda$ the diffusion of the cloud leads to smallness of the
part of $\Psi$ which passes the potential $U$, and $\Xi$ becomes closer to
unity. Thus, function $\Xi$ has long tail from potential with direction to
negative $\phi_1$ where it is not equal to $1$. The procedure of finding
$\Xi$ from the equation (\ref{zero}) corresponds to the geometrical optics
approximation (taking into account only derivatives in propagating direction;
this allows one to get the fact of propagation). This tail of $\Xi$ in this
approximation is nothing else than the shadow of potential $U$. Higher orders
of perturbation theory over the transverse derivatives correspond to
diffraction corrections.

Now let us consider the correction $\xi$. The equation for it looks
like $( \hat\Gamma_1 + \hat\gamma) \xi = -\hat\gamma \Xi_1$. Again
let us neglect $\hat\gamma$ in l.h.s. and solve the equation. $\Xi_1$
is some exponential function with scale of the order 1. Then
$\hat\gamma \Xi_1 \sim \Xi_1$. Not that $\hat\gamma \Xi_1$ is almost
equal to zero at $\phi_1 > \phi_\Lambda^+$. To estimate $\xi$ one
should construct Green function $G(\phi_1|\phi_0)$ for operator
$\hat\Gamma_1$:
\begin{equation}
  G(0|\phi_0) \simeq \frac1{2\lambda} \, {\rm exp} \left( - \left(
    \sqrt{\lambda^2 + \frac{y^2 P_2}{Dd}} - \lambda
    \right) \phi_0 \right) \left( 1 - C {\rm exp} \left( -2
    \sqrt{\lambda^2 + \frac{y^2 P_2}{Dd}} (\phi_\Lambda^+ -
    \phi_0) \right) \right) \,,
\label{ccc} \end{equation}
where
$$C =\left(\sqrt{\lambda^2 + y^2 P_2/Dd}-\lambda\right)/
\left(\sqrt{\lambda^2 + y^2 P_2/Dd}+\lambda\right) \,. $$ The unity in
brackets in (\ref{ccc}) gives the correction for $\Xi$ which has the same
exponential factor as $\Xi_1$. Thus $\xi$ does not change the answer with the
logarithmic accuracy. The second term in the brackets gains while $\phi_0$ is
close to $\phi_\Lambda^+$. It is due to nonzero width of the cloud $\Xi$ and
to dependence of $t_{\rm lt}$ on other variables.  Again, it does not change
the exponent.

The case of the passive scalar differences can be considered in a similar
way.

\end{document}